\documentclass[preprint,11pt]{aastex}
\usepackage{epsfig,natbib,emulateapj5}
\shortauthors{Eracleous et al.}
\shorttitle{Spectroscopy of UV-Bright Stars}

\slugcomment{To appear in the 2002 February issue of P.A.S.P.}
\begin{document}

\def\aj{\rm{AJ}}                   
\def\araa{\rm{ARA\&A}}             
\def\apj{\rm {ApJ}}                
\def\apjl{\rm{ApJ}}                
\def\apjs{\rm{ApJS}}               
\def\apss{\rm{Ap\&SS}}             
\def\aap{\rm{A\&A}}                
\def\aapr{\rm{A\&A~Rev.}}          
\def\aaps{\rm{A\&AS}}              
\def\mnras{\rm{MNRAS}}             
\def\nat{\rm{Nature}}              
\def\pasj{\rm{PASJ}}    	   
\def\procspie{\rm{Proc.~SPIE}}     

\title{A Spectroscopic Reconnaissance of UV-Bright Stars\altaffilmark{1}}

\author{
Michael Eracleous\altaffilmark{2}, Richard A. Wade, \& Mala Mateen}
\affil
{Department of Astronomy and Astrophysics, The
Pennsylvania State University, 525 Davey Lab, University Park, PA  16802}
\email{mce@astro.psu.edu, wade@astro.psu.edu, mateen@astro.psu.edu}
\and 
\author{Howard H. Lanning}
\affil{Computer Sciences Corporation/Science Programs, Space 
Telescope Science Institute, 3700 San Martin Drive, Baltimore, MD 21218}
\email{lanning@stsci.edu}

\altaffiltext{1}{Based, in part, on observations with the {\it
Hobby-Eberly Telescope}.}

\altaffiltext{2}{Visiting Astronomer, Kitt Peak National Observatory, National
Optical Astronomy Observatory, which is operated by the Association of
Universities for Research in Astronomy, Inc. (AURA) under cooperative
agreement with the National Science Foundation.}

\begin{abstract}
We have carried out spectroscopic observations and made preliminary
classifications of 62 UV-bright stars identified by Lanning on plates
taken by A.\ Sandage. The goal was to search for ``interesting''
objects, such as emission-line stars, hot sub-dwarfs, and high-gravity
stars. Our targets were grouped into two samples, a bright ($m_{\rm
B}<13$) sample of 35 stars observed with the Kitt Peak 2.1m telescope
and a faint ($13<m_{\rm B}<16$) sample of 27 stars observed with the
{\it Hobby-Eberly Telescope}.  We find 39\% fairly normal O--mid~B
stars, 15\% late~B--late A stars and 32\% F--G stars, with 13\% of the
stars being high gravity objects, composite, or otherwise
peculiar. Included are four emission-line stars, three composite
systems. Thus one out of every ten Lanning stars is ``interesting''
and may deserve individual study.  Stars in the bright sample are
often found to be late F or early G stars, although this sample does
include interesting stars as well. No such large contamination occurs
among the fainter stars, however, owing to ``deselection'' of these
stars by interstellar reddening in the low-latitude fields of the
survey.

\end{abstract}

\section{Introduction}

Lanning has published six lists of UV-bright stars in the Galactic
Plane, based on eye inspection of two-color photographic plates
obtained by A.\ Sandage using the Palomar Schmidt telescope
\citep{p_I,p_II,p_III,p_IV,p_V,p_VI}.  Magnitudes range from about 10
to about 20, with $U-B$ colors typically bluer than $-0.2$.  As
demonstrated by \citet{md81} the Lanning lists are a potentially rich
source of rare or important objects, if the stars can be observed
spectroscopically.  Those authors reported spectroscopy of nineteen
objects drawn from list I, finding two emission line objects in the
group. Of these, one turned out to be a cataclysmic variable (Lanning
10 = V363 Aur) and the other is an emission-line shell star.  Other
Lanning stars have turned out to be previously known objects, such as
HZ~Her (Her~X-1), WZ~Sge (a recurrent nova), white dwarfs, central
stars of faint planetary nebulae (e.g., HFG~1), and various stars
named in the Luminous Stars of the Northern (or Southern) Milky Way
Survey (LS/LSS).

Since lists II--VI of the Lanning catalog have been published
relatively recently, such spectroscopy is not generally available in
the literature for most of the 459 objects. As noted above, some
Lanning stars have proposed cross-identifications with previously
known objects, based on positional and magnitude coincidence.
However, besides the objects examined by \citet{md81}, or followed up
by others after that work called attention to them, only Lanning 90 =
V1776~Cyg \citep{Shafter,Garn} and Lanning 159 \citep{LiuHu} have been
newly described spectroscopically.  Thus we have undertaken a
spectroscopic reconnaissance of a subset of Lanning stars, the results
of which are the subject of this paper. 

In \S{2} we describe the target selection, observations and data
reduction.  In \S{3} we present our classification of the spectra and
a census of interesting objects, which we discuss in more detail in
\S{4}. We close with a brief summary of our findings in \S{5}.

\section{Targets, Observations, and Data Reduction}

We selected two samples of stars from lists I, II and (mostly) V by
\citet{p_I} and \citet{p_II,p_V}, which we refer to hereafter as the
``bright'' and ``faint'' samples. This selection was dictated in part
by the timing of our observing runs. The stars lie along the northern
Milky Way between Right Ascensions 19h and 22h and in the declination
range $+18^{\circ}$ to $+62^{\circ}$. Updated coordinates of all
Lanning stars along with any proposed cross-identifications can be
found at {\tt http://www.stsci.edu/\~{}lanning/index.html}.  In
general, we have selected stars for spectroscopic follow-up which do
not have existing cross-identifications.  More than half of all the
targets and all of the members of the bright sample are from the 
recently published and therefore poorly studied list V \citep{p_V}. The
sample of stars that we observed is representative of the range of
$U-B$ colors of stars included in these lists, and it should be
representative of the types of stars to be found in the magnitude
range examined.

The bright sample includes 35 stars brighter than $m_{\rm B}\approx
13$, while the faint sample includes 27 stars with $m_{\rm
B}=13-16$. The members of the two samples are listed in
Tables~\ref{brighttargs} and \ref{fainttargs}.  Two of the stars of
the faint sample were observed more than once, with different
observations appearing as separate entries in Table~\ref{fainttargs}.

The bright sample stars were observed with the Kitt Peak National
Observatory's (KPNO) 2.1m telescope and GoldCam spectrograph on 2000
September 24--25 with exposures of 300~s. The observation epochs are
listed in Table~\ref{brighttargs}. We used a $1.\!\!^{\prime\prime}8$
slit and a 500~mm$^{-1}$ grating to achieve a spectral resolution of
4.9~\AA\ over the spectral range 3875--7530~\AA. The observations were
made either during small gaps in the primary observing schedule or
during marginal weather.

The observations of the faint sample were carried out in
queue-scheduled mode with the {\it Hobby-Eberly Telescope} ({\it HET})
and Marcario Low-Resolution Spectrograph (LRS) in 2000 May and July
with exposure times ranging between 120 and 1500~s. The specific
observation epochs as well as the exposure times are listed in
Table~\ref{fainttargs}.  We employed a $1^{\prime\prime}$ slit and a
600~mm$^{-1}$ grism, which yielded a spectral resolution of 4.5~\AA\
and covered the range 4260--7245~\AA.

The spectral resolution and spectral coverage used in the observations
obtained with the KPNO 2.1m telescope were determined by the
objectives of the unrelated, primary observing program.  In the case
of the {\it HET} observations, we chose the highest resolution mode
available with the LRS; note that the spectral coverage is determined
by the choice of grism.  The resulting resolution and wavelength
coverage are similar for the two instruments but differ from those
normally used in classifying stars on the MK system.  They are, however,
well-suited for our purpose of carrying out a reconnaissance of
Lanning stars, where the goals are to assess the distribution of
spectral types and to identify unusual objects for possible later
study.  The coverage of redder wavelengths (H$\alpha$, molecular
bands) is especially useful for the latter goal.

The spectra were reduced in a standard manner. In summary, following
bias subtraction and flat field division, we traced and extracted
spectra from a wide window along the slit, which included most of the
stellar flux, subtracting the contribution from the night sky.  An
initial wavelength scale was derived from spectra of arc lamps taken
either at the beginning of the night or interspersed among the target
observations. The root-mean square residuals of a polynomial fit to
the arc line wavelengths were less than 0.1 pixels. The zero point of
the wavelength scale was refined with the help of strong night-sky
emission lines that were recorded in the same exposure as the star,
yielding a final absolute scale good to 0.3--0.5~\AA.  The flux scale
was established with the help of observations of spectrophotometric
standard stars. Uncertainties in the relative flux scale are less than
10\% throughout most of the spectral range and increase to 15--17\% at
wavelengths longer than 6500~\AA. The absolute flux scale can be
uncertain by as much as a factor of 2, especially in the case of the
{\it HET} observations, since the spectra were taken through a narrow
slit to assure best resolution under variable seeing and fluctuating
image quality. Discrete atmospheric absorption bands were corrected
using templates derived from the spectra of the spectrophotometric
standards stars, which are featureless in the regions of interest.

\section{Classification and Census of Interesting or Unusual Objects}

In order to classify the program stars, we used the digital {\it
Library of Stellar Spectra} \citep*{jhc84} as a source of ``standard''
comparison spectra. This was possible since the resulting resolution
of our spectroscopy is very similar to theirs, and the overlap in
wavelength coverage is adequate.  

For classification purposes it was desirable to remove the major
effects of interstellar reddening from the observed spectra,
especially since the Jacoby et al.\ spectra were published with this
correction applied.  At the same time, it was desirable to retain some
of the undulations in the spectrum that help to indicate spectral
type.  We thus {\it approximately} normalized both the program and
standard spectra in the same way, using a broken straight line fit: we
divided the blue portion of each spectrum by a straight line connecting
the average flux density in a small ``blue'' wavelength interval to the
average flux density in a small ``yellow'' interval, and we divided the red
portion of each spectrum by a straight line fit connecting the same
yellow interval with the average flux density in a small ``red'' wavelength
interval.  Thus the spectra are pivoted about the center of the yellow
interval, which is at 5500~\AA. (The blue and red intervals are
centered at 4040~\AA\ and 7030~\AA, respectively, for the spectra
acquired at KPNO; and at 4410~\AA\ and 7030~\AA, respectively, for the
spectra acquired at the {\it HET}.  For comparison with a given
program star's spectrum, the Jacoby et al.\ standard spectra were
normalized with the appropriate choice of intervals.)  This
approximate normalization sufficed to remove the major effects of
interstellar reddening (which resembles a line broken at 5500~\AA) and
the bulk of the curvature of the intrinsic spectrum of each star,
leaving shorter-scale undulations along with the absorption line
spectrum of each star.  Our normalization scheme is simple, rapid, and
objective in execution, and entirely adequate for the purposes of our
present study.

A rough classification was then carried out by visual inspection,
comparing a subset of the spectra from Jacoby et al.\ with each
program star.  The subset of Jacoby et al.\ stars included stars of
spectral type O7, O9, B0, B3, B8, A2, A3, A7, F0, F6, G0, K0, K4, M0,
and M5, all of luminosity class V, along with two metal-weak stars of
types F4 and F7.  Normalized line depths and line depth ratios were
the principle diagnostics of spectral type. Additional information was
drawn from the shape of the continuum after normalization. The main
purpose of the classification was to divide the program stars into
early and late spectral types and to note unusual spectra, including
composite-spectrum and high-gravity stars. The entire observed
wavelength range was used initially.  For the bright sample, the
classification was repeated using expanded plots of the 3800--5300\AA\
region, and these classifications are preferred in
Table~\ref{brighttargs}; the two classifications were usually
consistent, in one case a dubious classification being clarified using
the expanded ``blue'' plots.  Further refinement in the spectral
types, perhaps especially with respect to luminosity class of the
earlier stars, may be possible with the present data, but such work
would be better pursued with data that have somewhat higher resolution
and additional coverage at shorter wavelengths.  Our estimated
spectral types are given in Tables~\ref{brighttargs} and
\ref{fainttargs}, along with brief notes where warranted.  In some
cases a range of possible spectral types is given. In other cases, the
best matching type from the set of standards is given; in this case a
tilde ($\sim$) preceding the type indicates a somewhat broader range
of allowed matches is possible but could not be explored with our
sparse selection of standards.  A colon (:) indicates a difficult
classification.

Of the 35 stars in the bright sample, we classify seven stars as O to
mid-B, nine stars as late-B to late-A, and nineteen stars as F or G.
This group includes a large number of late F to early G stars. These
are likely metal--poor or have large individual color errors, in order
to have a sufficiently extreme $U-B$ color for inclusion in the
catalog (see further discussion below). There are no obvious candidates
for high--gravity stars. Some of the stars are strongly reddened, as
shown by both the continuum shape before normalization and the
presence of diffuse interstellar bands. For a sample of fourteen
early-type stars an eye-estimate was made of the depth (in continuum
units) of the unresolved $\lambda$6283 interstellar band. There is a
positive but weak correlation between the band depths and $U-B$ color
excesses, which were estimated using $U-B$ from \citet{p_II,p_V} and
intrinsic colors inferred from the assigned spectral types. Such a
correlation is expected, since both interstellar absorption bands and
reddening are due to ``dust.''  Its weakness in the present sample is
largely owing to observational scatter in the inferred and measured
colors of the stars.

The fainter sample of 27 Lanning stars contains 16 stars classified by
us as O or B, three stars called ``hot, high gravity,'' five DA, DA +
late-type, or other composite stars, one F--G star, one He-strong
star, and a star surrounded by an emission nebula. In addition, we
observed the M dwarf G209--30, which is a high proper motion star in
the field of Lanning 93.  This sample shows the reverse properties of
the bright sample, namely it includes more high-gravity stars and only
one late F--G star. We interpret this as a result of color
``deselection'' of marginal stars by interstellar reddening.  F--G
stars are expected to have $U-B$ colors redder than the nominal cutoff
adopted by Lanning for generating the catalog from the Sandage plates.
They can, however, enter the Lanning sample if they are metal weak
(i.e., subdwarfs, with reduced line blocking in the ultraviolet) or if
the individual photometric errors are large; in either case, their
$U-B$ colors will still be close to the cutoff.  Fainter stars on
average will be more distant, hence more heavily reddened, so that
F--G stars among them no longer appear in a color-selected catalog,
even if they are subdwarfs or in the case of large photometric
errors. Likewise, many (but not all) luminous OB stars may be very
heavily reddened if they are apparently faint in these Milky Way
fields.  Thus fainter Lanning stars in the sample tend to be of lower
intrinsic luminosity, since that is what remains of the unreddened
nearby sample; or to have very negative intrinsic $U-B$ color, to
withstand significant reddening.

Several unusual spectra, including composite hot+cool systems,
emission line stars, and He-strong stars were found. In particular,
Lanning 121, Lanning 386 and Lanning 441 are composite, with \ion{Mg}{1}~b,
\ion{Na}{1}~D, TiO and other absorption features indicative of a
late-type star apparent in the spectra, along with H or \ion{He}{1}
lines from a hotter star.  Lanning 388 is ``He-strong,'' that is,
moderately strong \ion{He}{1} and \ion{He}{2} lines are present along
with strong H Balmer lines; \ion{He}{1} $\lambda$4922, for example, is
nearly as deep as H$\beta$, unlike in the stars from the Jacoby et
al.\ atlas to which it was compared (HD~35619 O7~V, HD~12323 O9~V, 
HD~158659 B0~V).

Several of the 62 Lanning stars sampled are thus ``interesting'', not
counting DA white dwarfs and other possible high-gravity stars.  In
the following section we discuss the properties of some of these
objects further.

\section{Notes on Individual Objects}

For the emission-line stars discussed below, we attempted to use the
strength of the $\lambda 4430$ diffuse interstellar band (DIB) to
estimate the reddening and hence the extinction and distance.  We
prefer this method instead of comparing observed spectral energy
distributions (SEDs) and intrinsic SEDs inferred from the spectral
types, because uncertainties in the inferred spectral type are large
enough that the intrinsic SEDs of the stars cannot be accurately
determined.  Moreover, the spectra of these stars include emission
lines and therefore possibly a non-stellar continuum, which could
distort the observed SEDs.  We used the calibrations of the $\lambda
4430$ DIB as a reddening indicator reported by \citet{w63,w66},
\citet{h75}, and \citet{tsk81} and we have assumed a relationship
between visual extinction and distance of $A_{\rm V}/d = 0.8~{\rm
mag~kpc}^{-1}$ \citep{allen73}.  Using the resulting distances and
extinction-corrected fluxes, we are able to report approximate
emission-line luminosities. Uncertainties in reddening, assessed from
the dispersion of results obtained from different methods, are of
order 0.2~mag. Uncertainties in the absolute flux calibration due to
slit losses lead to error bars of a factor of 2 in quoted
luminosity. There is also an uncertainty in converting visual
extinction to distance, which is more difficult to assess in patchy
Milky Way fields, that we have not taken into account.


\noindent{\it Lanning 19 and Lanning 21. ---}
Very broad and rather shallow absorption lines of H are present in
these stars.  If present, \ion{He}{1} is very weak.

\noindent{\it Lanning 23. ---} Observed by
\citet{md81} who remarked, ``Balmer abs; late B?''.  \citet{f_etal96}
classified this star as DA1; it was detected in the ROSAT All-Sky
Survey.

\begin{figure}
\plotone{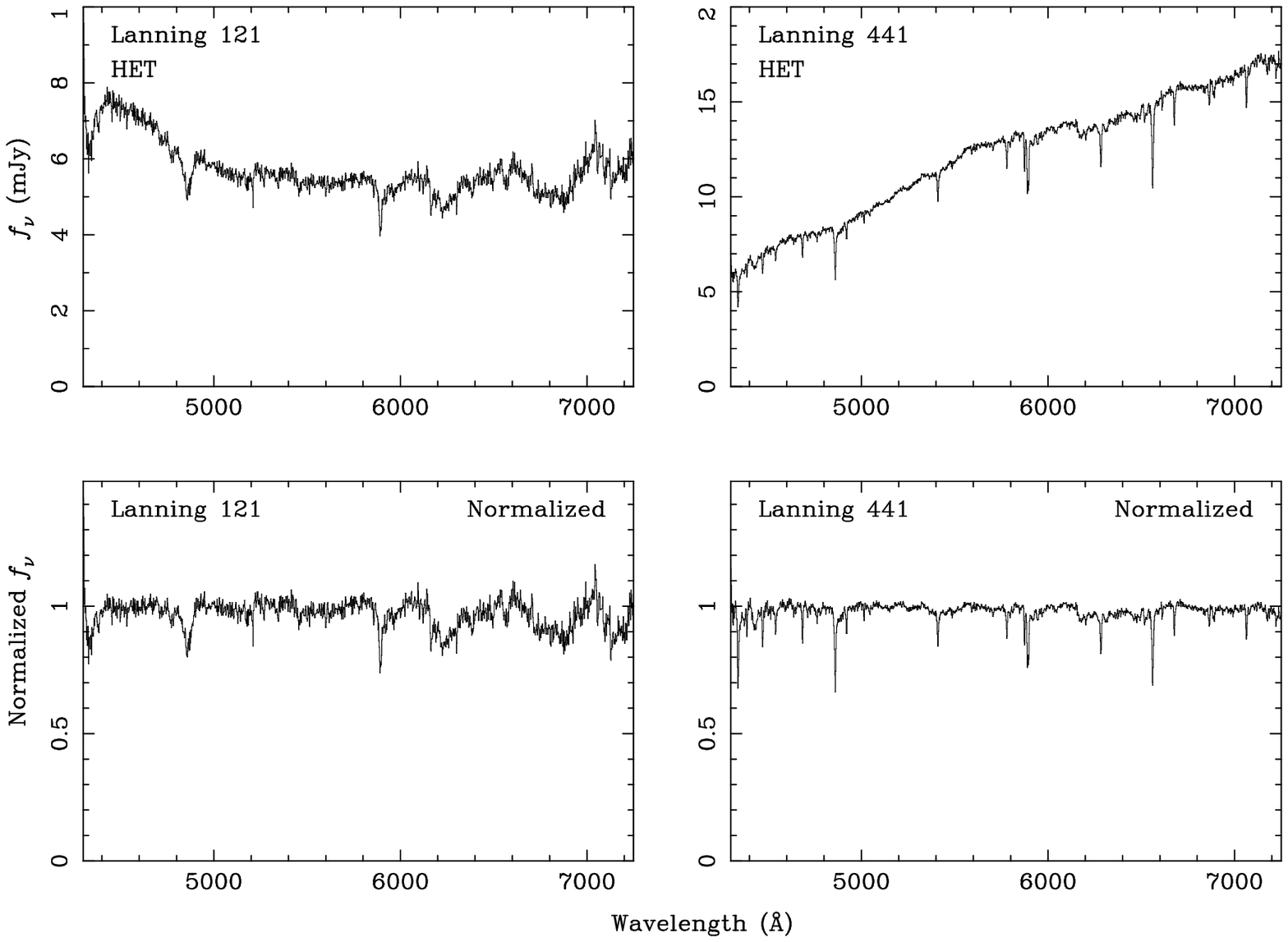}
\caption{\label{figabstars} Spectra of Lanning~121 and
Lanning~441 (both from the {\it HET}), showing their composite
nature. The top panels show the original spectra, while the 
bottom panels show the normalized spectra.
Their classification is given in Table~\ref{fainttargs} and
their properties are discussed in \S{4} of the text.}
\end{figure}

\noindent{\it Lanning 121. ---} Very broad
absorption lines of H, but not He, are present, along with atomic and
molecular band features indicating a late--K or M-type companion star.
The spectrum is shown in Figure~\ref{figabstars}.

\begin{figure}
\plotone{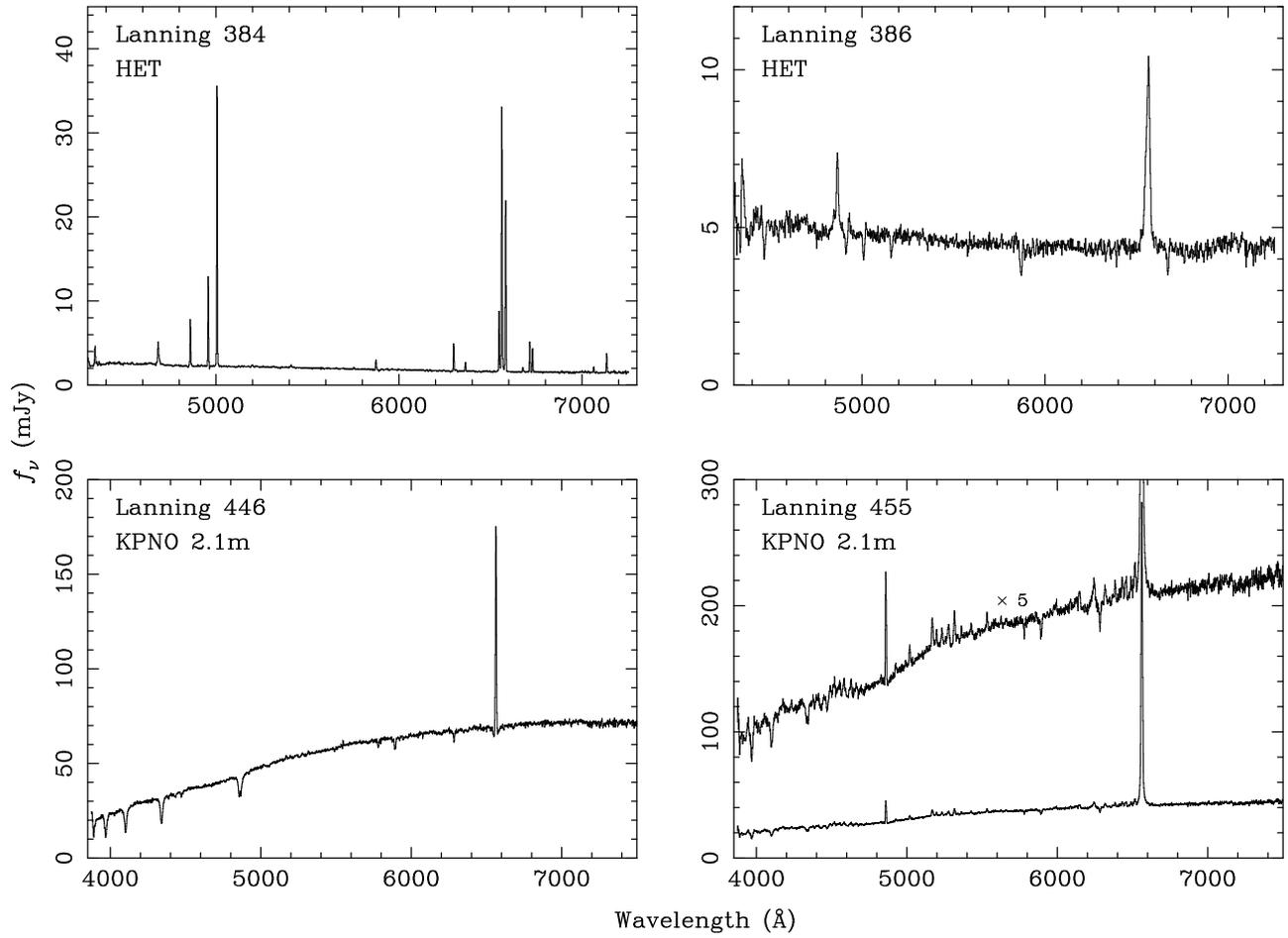}
\caption{\label{figemstars} Spectra of the four emission line stars
found.  The first two objects were observed with the {\it HET}, while
the last two were observed with the KPNO~2.1m. Their properties are
discussed in detail in \S{4} of the text and their classification is
given in Tables~\ref{brighttargs} and \ref{fainttargs}. In the case of
Lanning 455 we also show a magnified view of the spectrum (by a factor
of 5) to better illustrate the forest of \ion{Fe}{2} lines.}
\end{figure}

\noindent{\it Lanning 384. ---} This star, which is located in Cygnus,
is surrounded by an extended, emission-line nebula, whose spectrum
includes many strong permitted and forbidden lines from a wide range
of ionization states (from \ion{S}{2} to \ion{He}{2}). The spectrum of
the star itself is rather blue with no obvious absorption lines; some
filling in by emission may have occurred. The nebula is visible in the
red, second-generation Palomar Observatory Sky Survey plate of this
region.  It is elliptical in shape with a major axis of length
$4^{\prime}$, oriented approximately E--W. The inner part of the
nebula is particularly bright and appears to define an inner ellipse
whose major axis is approximately at right angles to that of the outer
nebula. By chance, the spectrograph slit was set at a position angle
of $270^{\circ}$, which is along the major axis of the outer
nebula. The sky-subtracted 2-dimensional spectrum shows the
low-ionization emission lines ([\ion{S}{2}]~$\lambda\lambda$6717,6731,
[\ion{N}{2}]~$\lambda\lambda$6548,6583, Balmer lines) extending along
the entire length of the slit of $4^{\prime}$. There are knots of
enhanced emission in the strongest low-ionization lines at a distance
of about $45^{\prime\prime}$ from the central star, which also appears
to be the extent of the \ion{He}{2}-emitting region of the nebula. The
spectrum of the innermost parts of the nebula, extracted from a window
of width $10^{\prime\prime}$, is shown in Figure~\ref{figemstars}, and
the relative emission-line strengths measured from it are reported in
Table~\ref{emlines}, with no extinction corrections applied. Since the
slit was oriented at the parallactic angle during this observation,
the relative line strengths should not be distorted by differential
atmospheric refraction at the slit.  From the relative strengths of
the emission lines we can infer some of the basic properties of the
inner nebula. If we require that the intrinsic H$\alpha$/H$\beta$
ratio follows case~B recombination, we infer a reddening of
$E(B-V)=0.2$. From the ratio of [\ion{O}{3}]~$\lambda\lambda4959,5007$
to [\ion{O}{3}]~$\lambda4363$, we estimate a temperature of 15,000~K,
while from the ratio of the lines in the
[\ion{S}{2}]~$\lambda\lambda6717,6731$ doublet we estimate an electron
density of 120~cm$^{-3}$ \citep[see][]{osterbrock89}.  The mean
heliocentric velocity of this nebular gas is $-$40 km~s$^{-1}$,
measured from eleven permitted and forbidden transitions.

\noindent{\it Lanning 386. ---} The spectrum
of this star sports a blue continuum with \ion{He}{1} absorption lines
as well as late-K or M features plus broad Balmer emission lines. From
the strength of the $\lambda$4430 DIB we estimate $A_{\rm V}=1.5\pm
0.2$, hence a distance of $1.9\pm 0.2$~kpc and an H$\alpha$ luminosity
of $1.3\times 10^{32}~{\rm erg~s}^{-1}$. The H$\alpha$/H$\beta$ ratio
is 1.4 and the lines are rather broad: the H$\beta$ line has a FWHM
of 820~km~s$^{-1}$ while the H$\alpha$ line has a FWHM of
1140~km~s$^{-1}$ (corrected for instrumental broadening). The flat
Balmer decrement together with the emission-line luminosities and
widths suggests that this is a cataclysmic variable or related
interacting binary.  The spectrum is shown in Figure~\ref{figemstars}.

\noindent{\it Lanning 388. ---} An OB
star with strong \ion{He}{1} and \ion{He}{2} lines (see previous section).

\noindent{\it Lanning 441. ---} Absorption
lines of H, \ion{He}{1} and \ion{He}{2} are present, along with band
heads indicating an M-type companion star.  The spectrum is shown in
Figure~\ref{figabstars}.

\noindent{\it Lanning 446. ---} This Be star
has H$\alpha$ in emission, with the H$\beta$ absorption line partly
filled by emission. The continuum appears reddened by interstellar
and/or circumstellar absorption and the strength of the $\lambda$4430
DIB implies $A_{\rm V}=0.9\pm 0.2$ leading to a distance of $1.1\pm
0.3$~kpc and an H$\alpha$ luminosity of $2\times 10^{33}~{\rm
erg~s}^{-1}$.  The H$\alpha$ line is unresolved (down to a limiting
FWHM of 230~km~s$^{-1}$), which is consistent with an origin in
circumstellar matter.  The spectrum is shown in
Figure~\ref{figemstars}.

\noindent{\it Lanning 447. ---} Has shallow
and possibly broad absorption lines, indicating it may be a sdO star.

\noindent{\it Lanning 455. ---} This star shows both \ion{Fe}{2} and
Balmer emission lines. It appears as No.\ 469 in a list of H$\alpha$
emission stars compiled by \citet{mebu}, and is also known as
LS~III~$+55^{\circ}12$. The continuum appears to be reddened and the
$\lambda$4430 DIB implies $A_{\rm V}=2.3\pm 0.2$ leading to a distance
of $2.8\pm 0.2$~kpc and an H$\alpha$ luminosity of $1\times
10^{34}~{\rm erg~s}^{-1}$.  The H$\alpha$/H$\beta$ ratio is 6 (after
reddening correction) and the FWHM of the H$\alpha$ line is
430~km~s$^{-1}$ (corrected for instrumental broadening). The
equivalent widths of the H$\alpha$ and H$\beta$ emission lines are
4.7~\AA\ and 74.1~\AA, respectively. The spectrum resembles those of
T~Tauri stars in the collection of \citet{kc79}, suggesting that this
object may be a T~Tauri star as well.  If so, the reddening and
distance quoted combined with an apparent visual magnitude of
$\approx12$ would imply a very luminous, hence young object.

\section{Conclusions}

We have obtained spectra and made preliminary classifications of 62
UV--bright stars.  Of these 62 objects, four are emission-line stars,
three are composite systems (double-counting one emission-line star),
and another shows strong \ion{He}{1} absorption lines. From the
perspective of spectral classification, we find 39\% more or less
normal O--mid~B stars, 15\% late~B--late A stars and 32\% F--G stars,
with 13\% of the stars being high gravity objects, composite, or
otherwise peculiar.  Thus we reaffirm and greatly extend the
conclusion to be inferred from \citet{md81} that at least one out of
every ten Lanning stars is ``interesting'' and worthy of individual
study.

Lanning stars brighter than $m_B = 13$ are often found to be late F or
early G stars, at least in the fields described in paper V
\citep{p_V}.  No such large contamination occurs among the fainter
stars, however, owing to ``deselection'' of these stars by
interstellar reddening in the low-latitude fields of the
survey. Interesting stars nevertheless do appear in the bright sample.

\acknowledgements

This work was based in part on observations obtained with the {\it
Hobby-Eberly Telescope}, which is a joint project of the University of
Texas at Austin, Pennsylvania State University, Stanford University,
Ludwig-Maximillians-Universit\"at M\"unchen, and
Georg-August-Universit\"at G\"ottingen. The Marcario Low-Resolution
Spectrograph (LRS) is a joint project of the Hobby-Eberly Telescope
partnership and the Instituto de Astronom\'ia de la Universidad
Nacional Aut\'onoma de M\'exico.

The original photographic survey by Sandage was supported in part by
the National Aeronautics and Space Administration under grant NGR
09-140-009.  NASA support to Lanning has been from the Astrophysics
Data Program through contract PO\# S-92513-Z.  We have made use of the
SIMBAD database operated at CDS, Strasbourg, France, and NASA's
Astrophysics Data System Bibliographic Services.

{}


\clearpage

\clearpage


\begin{deluxetable}{rclcll}
\tablenum{1}
\tablewidth{6.5in}
\tablecolumns{6}
\tablecaption{Bright Sample:  Targets, Observations, Classification\label{brighttargs}}
\tablehead{
\colhead{Lanning} &
\colhead{Approx.} &
\colhead{Observation Date} &
\colhead{Exposure} &
\colhead{Spectral} &
\colhead{}
\cr
\colhead{Number} &
\colhead{$m_{\rm B}$} &
\colhead{and Time (UT)} &
\colhead{Time (s)} &
\colhead{Class} &
\colhead{Notes}
}
\startdata
     365 & 10.2 &  2000 Sep 24 06:50:55  &   300 & B8  &    \cr 
     368 & 10.5 &  2000 Sep 25 07:12:45  &   300 & mid B  &    \cr 
     369 & 10.6 &  2000 Sep 24 06:57:30  &   300 & early G  &    \cr 
     370 & 10.7 &  2000 Sep 24 07:25:41  &   300 & mid F -- G0  &    \cr 
     371 & 10.5 &  2000 Sep 24 07:07:14  &   300 & $\sim$ F6 &    \cr 
     373 & 10.8 &  2000 Sep 25 07:24:41  &   300 & B8   &    \cr 
     374 & 10.8 &  2000 Sep 25 07:30:43  &   300 & $\sim$ F6  &    \cr 
     375 & 10.5 &  2000 Sep 25 07:18:45  &   300 & $\sim$ F6  &    \cr 

     378 & 10.7 &  2000 Sep 25 07:45:26  &   300 & early G &    \cr 
     379 & 10.8 &  2000 Sep 24 07:13:39  &   300 & G  &    \cr 
     380 & 10.6 &  2000 Sep 25 07:06:41  &   300 & late F -- G &    \cr 
     381 & 10.6 &  2000 Sep 24 07:19:46  &   300 & mid F &    \cr 

     390 & 11.0 &  2000 Sep 25 07:51:35  &   300 & mid F  &    \cr 
     393 & 11.3 &  2000 Sep 25 07:39:19  &   300 & $\sim$ F6  &    \cr 
     397 & 11.2 &  2000 Sep 24 08:52:17  &   300 & $\sim$ B3  &    \cr 
     398 & 10.8 &  2000 Sep 25 08:46:39  &   300 & $\sim$ F8  &    \cr 
     401 & 11.6 &  2000 Sep 25 07:58:17  &   300 & late F -- G &    \cr 

     404 & 10.8 &  2000 Sep 24 07:41:25  &   300 & mid--late B  &    \cr 
     405 & 10.8 &  2000 Sep 24 08:58:44  &   300 & $\sim$ B8  &    \cr 
     406 & 12.4 &  2000 Sep 25 06:58:25  &   300 & late F  &    \cr 
     408 & 12.2 &  2000 Sep 25 08:06:32  &   300 & O7   \cr 
     412 & 10.9 &  2000 Sep 25 08:40:28  &   300 & O9 -- B0  &    \cr 

     419 & 10.5 &  2000 Sep 24 08:45:56  &   300 & $\sim$ F6  &    \cr 
     421 & 10.8 &  2000 Sep 25 08:28:36  &   300 & late F -- G &    \cr 
     428 & 10.9 &  2000 Sep 25 08:34:36  &   300 & $\sim$ F6  &    \cr 
     436 & 10.8 &  2000 Sep 25 08:21:37  &   300 & $\sim$ B8  &    \cr 
     446 & 12.0 &  2000 Sep 24 08:30:27  &   300 & mid--late B  & H$\alpha$ em., weak H$\beta$ abs.   \cr 
     447 & 11.0 &  2000 Sep 24 08:37:01  &   300 & late O -- early B & shallow H lines   \cr 
     448 & 11.5 &  2000 Sep 24 07:54:34  &   300 & A0--A2  &    \cr 
     449 & 12.2 &  2000 Sep 24 07:47:49  &   300 & A7  &    \cr 
     450 & 11.0 &  2000 Sep 25 08:15:37  &   300 & late F -- G0  &    \cr 
     452 & 11.5 &  2000 Sep 24 08:08:05  &   300 & late B  &    \cr 
     453 & 11.3 &  2000 Sep 24 08:00:38  &   300 & $\sim$ B3  &   \cr 
     454 & 11.4 &  2000 Sep 24 08:14:31  &   300 & G  &    \cr 
     455 & 12.0 &  2000 Sep 24 07:32:18  &   300 & O -- early B:  & H and \ion{Fe}{2} emission  \cr 
\enddata
\end{deluxetable}

\clearpage

\begin{deluxetable}{rclcll}
\tablenum{2}
\tablewidth{6.5in}
\tablecolumns{6}
\tablecaption{Faint Sample: Targets, Observations, Classification\label{fainttargs}}
\tablehead{
\colhead{Lanning} &
\colhead{Approx.} &
\colhead{Observation Date} &
\colhead{Exposure} &
\colhead{Spectral} &
\colhead{}
\cr
\colhead{Number} &
\colhead{$m_{\rm B}$} &
\colhead{and Time (UT)} &
\colhead{Time (s)} &
\colhead{Class} &
\colhead{Notes}
}
\startdata
      19 & 15.5 &  2000 Jul 23 09:11:46  &   320 & ---  &  hot, high-$g$? \cr 
      21 & 16.0 &  2000 Jul 08 08:29:00  &   240 & ---  &  hot, high-$g$? \cr 
         &      &  2000 Jul 10 10:06:14  &   620 &   &   \cr 
         &      &  2000 Jul 23 09:30:56  &   620 &   &   \cr 
      22 & 14.7 &  2000 Jul 24 09:37:10  &   130 & O7--O9  &   \cr 
      23 & 14.0 &  2000 Jul 23 08:18:40  &   120 & ---  &  v. hot, high-$g$ \cr 
      44 & 13.0 &  2000 Jul 24 08:55:17  &   120 & F8--G  &   \cr 
      45 & 13.5 &  2000 Jul 01 10:29:40  &   120 & mid-B  & \cr 
      49 & 14.5 &  2000 Jul 01 06:08:04  &   240 & O  &   \cr 
      80 & 14.5 &  2000 Jul 24 08:40:45  &   130 & $\sim$B5  &   \cr 
     105 & 16.0 &  2000 May 04 10:31:33  &   520 & $\sim$B3  &  \cr 
     106 & 13.0 &  2000 May 04 10:18:11  &   120 & $\sim$B3  &   \cr 
     110 & 15.0 &  2000 May 04 10:49:48  &   200 & mid-B  &   \cr 
     121 & 15.0 &  2000 Jul 08 07:02:51  &   600 & DA+late K--M  & composite  \cr 
     354 & 15.5 &  2000 Jul 04 06:07:47  &   420 & O--early B  &   \cr 
     367 & 15.5 &  2000 Jul 08 06:23:00  &   540 & B0--B3  &   \cr 
     382 & 16.0 &  2000 Jul 04 06:34:10  &   600 & mid-B  &   \cr 
     384 & 16.5 &  2000 Jul 31 10:24:18  &  1000 & ---  & emission nebula  \cr 
     385 & 16.0 &  2000 May 05 10:38:30  &   720 & mid-B  &   \cr 
     386 & 15.8 &  2000 Jul 10 06:03:23  &   430 & \ion{He}{1} abs.+late K--M0 & broad H em.; comp. \cr 
     388 & 16.5 &  2000 Jul 31 09:56:15  &  1000 & O  & He--strong   \cr 
     391 & 16.2 &  2000 May 05 10:11:24  &   900 & mid-B  &   \cr 
     399 & 13.8 &  2000 Jul 10 08:19:54  &   120 & $\sim$B3  &   \cr 
     411 & 16.8 &  2000 Jul 25 09:33:18  &  1080 & DA  &   \cr 
     423 & 16.5 &  2000 Jul 25 10:16:31  &   620 & early-B  &   \cr 
     437 & 15.5 &  2000 Jul 23 10:38:06  &   270 & mid--late B  &   \cr 
     441 & 14.5 &  2000 Jul 22 10:34:32  &   180 & sdO+dM? &  composite \cr 
     443 & 10.8 &  2000 Jul 31 09:28:30  &   600 & late B:  & noisy  \cr 
     459 & 16.0 &  2000 Jul 29 07:15:11  &  1200 & DA  &   \cr 
         &      &  2000 Jul 30 09:48:54  &  1500 &     &   \cr 
 G209-30 &      &  2000 Jul 31 09:10:37  &   300 & dM0--dM5  &   \cr 
\enddata
\end{deluxetable}


\begin{deluxetable}{lrclr}
\tablenum{3}
\tablewidth{4.0in}
\tablecolumns{5}
\tablecaption{Observed Emission Line Strengths of Lanning 384 \label{emlines}}
\tablehead{
\colhead{Ion and} & 
\colhead{} & 
\colhead{\phantom{MMM}} & 
\colhead{Ion and} &
\colhead{} \cr
\colhead{Transition} & 
\colhead{$F/F_{\rm H\beta}$} & 
\colhead{\phantom{MMM}} & 
\colhead{Transition} &
\colhead{$F/F_{\rm H\beta}$} 
}
\startdata
\ion{H}{1}    $\lambda$4341 & 0.53 & & [\ion{O}{1}] $\lambda$6300 & 0.34 \cr
[\ion{O}{3}]  $\lambda$4363 & 0.10 & & [\ion{O}{1}] $\lambda$6363 & 0.11 \cr
\ion{He}{2}   $\lambda$4686 & 0.57 & & [\ion{N}{2}] $\lambda$6548 & 0.68 \cr
\ion{H}{1}    $\lambda$4861 & 1.00 & & \ion{H}{1}   $\lambda$6563 & 3.54 \cr
[\ion{O}{3}]  $\lambda$4959 & 1.67 & & [\ion{N}{2}] $\lambda$6583 & 2.13 \cr
[\ion{O}{3}]  $\lambda$5007 & 5.15 & & \ion{He}{1}  $\lambda$6678 & 0.42 \cr
[\ion{N}{1}]  $\lambda$5198 & 0.09 & & [\ion{S}{2}] $\lambda$6717 & 3.30 \cr
[\ion{He}{2}] $\lambda$5412 & 0.04 & & [\ion{S}{2}] $\lambda$6731 & 2.63 \cr
[\ion{He}{1}] $\lambda$5876 & 0.19 & & \ion{He}{1}  $\lambda$7064 & 1.69 \cr
\enddata
\end{deluxetable}

\end{document}